# Shortcuts to adiabatic non-Abelian braiding on silicon photonic chips

Wange Song,[1,2,†] Xuanyu Liu,[1,†] Jiacheng Sun,[1,†] Oubo You,[2] Shengjie Wu,[1] Chen Chen,[1] Shining Zhu,[1] Tao Li,[1,*] and Shuang Zhang,[2,3,4,*]

[1]*National Laboratory of Solid State Microstructures, Key Laboratory of Intelligent Optical Sensing and Manipulations, Jiangsu Key Laboratory of Artificial Functional Materials, College of Engineering and Applied Sciences, Nanjing University, Nanjing, 210093, China.*

[2]*New Cornerstone Science Laboratory, Department of Physics, University of Hong Kong, Hong Kong, China.*

[3]*Department of Electronic and Electrical Engineering, University of Hong Kong, Hong Kong, China.*

[4]*Materials Innovation Institute for Life Sciences and Energy (MILES), HKU-SIRI, Shenzhen, China.*

[†]These authors contributed equally.

*Corresponding authors: taoli@nju.edu.cn, shuzhang@hku.hk

The non-Abelian braiding describes the exchange behavior of anyons, which can be leveraged to encode qubits for quantum computing. Recently, this concept has been realized in classical photonic and acoustic systems. However, these implementations are constrained by adiabatic conditions, necessitating long operation distances and impeding practical applications. Here, we conceive and demonstrate a shortcut to adiabatic (STA) braiding of telecommunication light in three-dimensional silicon photonic chips. Our device comprises tri-layer silicon waveguides stacked and embedded in the SU-8 polymer, employing an STA strategy to expedite the braiding operations and give rise to compact devices that function as photonic quantum X, Y, and Z gates. We further experimentally observed non-Abelian braiding behaviors based on this STA-braiding scheme. Remarkably, this achievement represents the most compact braiding apparatus ever reported, with a size reduction of nearly three orders of magnitude compared to previous works. This work presents a feasible approach to accelerating adiabatic braiding evolutions, paving the way for compact, CMOS-compatible non-Abelian photonic devices.

Non-Abelian phenomena are pervasive and have been extensively explored across various fields of physics, including high-energy physics, condensed matter physics, and classical wave systems such as light and sound[1,2]. Despite their diversity, noncommutativity lies at the heart of non-Abelian phenomena, rendering the physics of non-Abelian systems significantly more complex and diverse than that of their Abelian counterparts. Examples include the non-Abelian gauge field used to explain the strong interaction[3], non-Abelian anyons and their statistics for describing the celebrated fractional quantum Hall effect[4-8], and the non-Abelian topological charges in braiding topological structures with multiple entangled bandgaps[9-11]. Notably, the presence of non-Abelian anyons in two-dimensional condensed matter systems has garnered increasing interest[4-6,12-14]. When non-Abelian anyons are swapped by intertwining them along world lines, their wavefunction exchange behavior is represented by a unitary matrix fundamentally different from that of exchanging fermions or bosons[15]. These anyons can be encoded with qubits to achieve quantum logic and fault-tolerant topological quantum computing. However, their implementation in quantum systems often depends on the dynamic winding of anyons, which can be challenging to achieve[16-19].

Interestingly, the non-Abelian braiding of degenerate zero modes has been extended to classical wave systems using light and sound as platforms[20-23], emerging as a multi-mode geometric effect known as the Berry–Wilczek–Zee (BWZ) phase[24,25], a matrix generalization of the well-known scalar Berry phase. The braiding of multiple states with non-Abelian characteristics, such as the Thouless pumping of flat-band modes, has also been successfully realized in photonics and acoustics[26-28]. While the associated non-commutative operations hold promises for applications involving unitary matrices, such as photonic quantum logic, these braiding operations are fundamentally restricted to adiabatic conditions, which require a sufficiently long distance and thus hinder practical applications. Consequently, compact photonic non-Abelian systems are highly sought after for investigating more intricate non-Abelian phenomena and developing practical applications in photon and light manipulations.

In this work, we theoretically propose and experimentally demonstrate a shortcut to adiabatic (STA) non-Abelian braiding of photonic modes in 3D silicon photonic chips. The braiding is achieved by imposing cyclic modulation of the hopping amplitudes among integrated silicon waveguides arranged in a trilayer configuration. An STA strategy[29-35] is proposed to identify an evolution pathway in the parameter space of braiding, enabling fast braiding operations and compact devices functioning as Pauli X, Y, and Z gates. Furthermore, by varying the sequence of two distinct STA braiding processes involving three modes, we experimentally observed different outcomes from identical initial states—a hallmark of non-Abelian braiding. Notably, the STA braiding scheme enables the braiding of telecommunication light within a miniaturized footprint of 1.7×24 μm² per device unit, nearly three orders of magnitude shorter compared to the state-of-the-art laser-written photonic waveguide system (millimeter scale)[22]. This work illustrates the potential to explore non-Abelian physics using a fully integrated CMOS-compatible silicon platform, ensuring versatile on-chip light manipulations and paving the way for compact non-Abelian photonic integrated devices.

**Results**

**Model**

We first illustrate the photonic braiding model with two degenerate zero modes. Figure 1a presents the schematics of the STA braiding structure in a tri-layered silicon photonic chip, which comprises three main waveguides (A, B, and S) and an auxiliary waveguide X. The bottom layer includes waveguides A and B (red waveguides) positioned on a sapphire substrate, while the top layer consists of waveguide S (blue waveguide). Waveguide X consists of three disjoint sections as required by the braiding process (to be detailed below). The first and third sections are in the middle layer (green waveguides), whereas the second section (red waveguide) is located in the bottom layer. Waveguides A, B, and S are coupled to waveguide X through evanescent wave coupling $\kappa_{iX}$ ($i$ = A, B, and S), as indicated by the dashed lines in the cross-sectional view shown in Fig. 1b. The dynamics of photon propagation in the waveguide structure can be described by the following Hamiltonian[22,23]:

$$H(z) = \begin{pmatrix} \beta_X & \kappa_{AX}(z) & \kappa_{BX}(z) & \kappa_{SX}(z) \\ \kappa_{AX}(z) & \beta_A & 0 & 0 \\ \kappa_{BX}(z) & 0 & \beta_B & 0 \\ \kappa_{SX}(z) & 0 & 0 & \beta_S \end{pmatrix}, \quad (1)$$

where $\beta_{X,A,B,S} = \beta_0$ represents the propagation constant of waveguides and the coupling coefficient $\kappa_{iX}$ are set to be real and are specific functions of parameter $z$ (the propagation distance along the waveguides, i.e., the braiding direction, the waveguide length is $L$). Since all other waveguides must couple through waveguide X, the system can be divided into two groups: one consisting of waveguide X and the other comprising waveguides A, B and S. Consequently, the Hamiltonian Eq. (1) can be simplified as $H = \begin{pmatrix} 0_1 & \kappa^T \\ \kappa & 0_3 \end{pmatrix}$, where $0_3$ is an $3 \times 3$ zero matrix and $\kappa = (\kappa_{AX}, \kappa_{BX}, \kappa_{SX})^T$ represents the vector space of the three coupling parameters. This Hamiltonian clearly exhibits sublattice symmetry, which ensures the presence of $|3 - 1| = 2$ degenerate zero modes in this case (note that there are also two splitting modes, which are not the focus of this work). These two zero modes reside within the group of waveguides A, B, and S, forming the braiding subspace. As illustrated in Fig. 1c, the unit hopping vector $\hat{\kappa} = \kappa/|\kappa|$ defines a unit 2-sphere, with the two zero modes spanning a tangent plane on this 2-sphere (marked by the red and blue arrows in Fig. 1c). The braiding process consists of three steps (step 0-1, 1-2, and 2-3) and forms a cyclic modulation of $z$, driving the tangent vectors of zero modes in a holonomic parallel transport. This process encloses a solid angle of $\pi/2$ on the 2-sphere and realizes the Y gate $Y = \begin{pmatrix} 0 & -1 \\ 1 & 0 \end{pmatrix}$, a fundamental quantum logic components (U(2) operation) that swaps the two modes as $|\Psi_1\rangle \rightarrow |\Psi_2\rangle$ and $|\Psi_2\rangle \rightarrow -|\Psi_1\rangle$.

Specifically, the system's initial configuration ($z=0$) is located at the north pole of the parameter space, and it contains two zero modes, occupying waveguides A and B, respectively, i.e., $|\Psi_1(0)\rangle = (0,1,0,0)^T$, $|\Psi_2(0)\rangle = (0,0,1,0)^T$. Here, $|\Psi_i\rangle$ represents a state vector with its elements indicating the wavefunction in waveguides X, A, B, and S, respectively, i.e., $|\Psi_i\rangle=(\varphi_{Xi}, \varphi_{Ai}, \varphi_{Bi}, \varphi_{Si})^T$. The braiding process consists of three steps, each traversing a distance of $L/3$, as shown in Fig. 1c. In **step 0-1** represented by the black arc connecting point 0 and 1, $\kappa_{SX}$ ($\kappa_{AX}$) decreases (increases) from its maximum (zero) to zero (its maximum), while $\kappa_{BX}$ is kept at zero. The initial state $|\Psi_1(0)\rangle = (0,1,0,0)^T$ becomes $|\Psi_1(L/3)\rangle=(0, 0, 0, -1)^T$. The zero mode evolutions can be mapped onto a Bloch sphere, where the eigenstates evolve from the north pole to the south pole of the Bloch sphere formed by the subspace ($\varphi_A, \varphi_S$), acquiring a $\pi$ phase during the evolution (Fig. 1e, left panel). On the other hand, the initial state in waveguide B $|\Psi_2(0)\rangle$ remains the same, $|\Psi_2(L/3)\rangle= |\Psi_2(0)\rangle = (0,0,1,0)^T$, as $\kappa_{BX} =0$ during the step 0-1. In **step 1-2**, the state in waveguide B is transferred to site A with a geometric phase $\pi$, i.e., from $|\Psi_2(L/3)\rangle= (0,0,1,0)^T$ to $|\Psi_2(2L/3)\rangle= (0,-1,0,0)^T$. This corresponds to the zero mode evolution from the south pole to the north pole of the Bloch sphere formed by subspace ($\varphi_A, \varphi_B$) (Fig. 1e, middle panel). In contrast, the state in waveguide S remains unchanged $|\Psi_1(2L/3)\rangle= |\Psi_1(L/3)\rangle = (0, 0, 0, -1)^T$ as $\kappa_{SX} =0$. In **step 2-3**, the state in waveguide S transfers to the waveguide B and also acquires the $\pi$ phase $|\Psi_1(L)\rangle= (0,0,1,0)^T$. Accordingly, the eigenstates evolve from the north pole to the south pole of the Bloch sphere formed by the subspace ($\varphi_S, \varphi_B$) (Fig. 1f, right panel). On the contrary, the state in waveguide A remains unchanged, i.e., $|\Psi_2(L)\rangle=|\Psi_2(2L/3)\rangle= (0,-1,0,0)^T$ as $\kappa_{AX} =0$. Overall, the initial states $|\Psi_1(0)\rangle=(0,1,0,0)^T$ and $|\Psi_2(0)\rangle=(0,0,1,0)^T$ finally become $|\Psi_1(L)\rangle=(0,0,1,0)^T$ and $|\Psi_2(L)\rangle=(0,-1,0,0)^T$ (marked by the two pentagrams in Fig. 1e), completing the braiding operations and functioning as quantum Y gates.

Notably, the braiding process typically requires adiabatic conditions[20-23,26-28], which necessitate a long operation distance. If the evolutionary distance is simply reduced, the system parameters will vary rapidly, leading to instantaneous eigenmode coupling and consequently undermining the braiding function. In order to address this issue, we propose to use shortcuts to adiabaticity (STA) braiding strategy to accelerate the adiabatic process, thus ensuring the desired state evolutions with a significantly reduced system length, as will be elaborated below.

**Shortcut to adiabatic (STA) braiding**

Transitionless driving, a well-known STA method proposed by M. Berry[29], can inhibit immediate eigenmode coupling caused by rapid parameter variations by incorporating *counterdiabatic driving* terms[30-35]. Since the fundamental component of the braiding process is the adiabatic zero-mode evolution within a triplet subsystem, we apply the STA strategy to this basic process and then extend it to the entire braiding procedure. Specifically, we consider the first step (0-1) of the braiding process as a subsystem, as indicated by the dashed box in Fig. 2a, which only involves waveguides A, X, and S. The triplet unit supports a zero mode $|\varphi_0\rangle$ and two split modes $|\varphi_\pm\rangle$, with the corresponding Hamiltonian (in the subspace of $(\varphi_A, \varphi_X, \varphi_S)$):

$$H(z) = \begin{bmatrix} \beta_A & \kappa_{AX}(z) & 0 \\ \kappa_{AX}(z) & \beta_X & \kappa_{SX}(z) \\ 0 & \kappa_{SX}(z) & \beta_S \end{bmatrix}. \tag{2}$$

Before applying the STA operation to Eq. (2), we note that this Hamiltonian satisfies the *one-photon resonance* condition (i.e., $\beta_A=\beta_X=\beta_S$), allowing us to map $H(z)$ to an effective two-level system $H_{eff}(z) = \begin{bmatrix} \kappa_{SX}(z)/2 & \kappa_{AX}(z)/2 \\ \kappa_{AX}(z)/2 & -\kappa_{SX}(z)/2 \end{bmatrix}$[31,34,35]. This pre-mapping prevents the generation of complex or long-range couplings during the STA operations[31], thereby enabling a feasible STA braiding system.

Following Berry's protocol[29], the counterdiabatic driving term $H_{eff\_CD}(z)$ of the mapped system has the form:

$$H_{eff\_CD}(z) = i \sum_n \sum_{m \neq n} \frac{|m(z)\rangle\langle m(z)| \frac{\partial}{\partial z} H_{eff}(z) |n(z)\rangle\langle n(z)|}{E_n(z) - E_m(z)}, \tag{3}$$

where $|m\rangle$ and $|n\rangle$ respectively denote eigenstates of the two-level systems with eigenvalues $E_m$ and $E_n$. Then the STA Hamiltonian of the effective system can be obtained by adding $H_{eff\_CD}(z)$ to $H_{eff}(z)$, i.e., $H_{eff\_STA}(z) = H_{eff}(z) + H_{eff\_CD}(z) = \begin{bmatrix} \tilde{\kappa}_{SX}(z)/2 & \tilde{\kappa}_{AX}(z)/2 \\ \tilde{\kappa}_{AX}(z)/2 & -\tilde{\kappa}_{SX}(z)/2 \end{bmatrix}$, where $H_{eff\_CD}(z)$ can inhibit the diabatic eigenmode transfer caused by rapidly changing $H_{eff}(z)$. Finally, the STA version of the original triplet $H_{STA}(z)$ is obtained by mapping $H_{eff\_STA}(z)$ back to the triplet

subsystem:

$$H_{STA}(z) = \begin{bmatrix} \beta_A & \tilde{\kappa}_{AX}(z) & 0 \\ \tilde{\kappa}_{AX}(z) & \beta_X & \tilde{\kappa}_{SX}(z) \\ 0 & \tilde{\kappa}_{SX}(z) & \beta_S \end{bmatrix}. \tag{4}$$

We see that the STA method modifies the coupling coefficients in the braiding process, which consequently alters the loop path along the braiding direction on the 2-sphere (see blue path in Fig. 1c). This STA Hamiltonian can be achieved by varying waveguides A, S, and X along the propagation direction (see schematics in Fig. 1a). The detailed procedure for performing STA braiding can be found in **Supplementary Information S1**.

The coupling strength of the conventional adiabatic design (following the stimulated Raman adiabatic passage (STIRAP)[36-38]) and the STA braiding process are shown in the top and bottom panels of Fig. 2a, respectively. For the light input from waveguide A, the transferred light intensity to waveguide S $|\varphi_S|^2$ as a function of the length of the subsystem (*step 0-1*) is shown in Fig. 2b. It is evident that the STA design offers the capability to complete the braiding process within a significantly reduced evolution distance; the minimum length required for the conventional pumping process to achieve complete transfer is 92 μm, while it could be shortened to 24 μm (approximately 3.8 times smaller) for the STA braiding. Specifically, the light evolutions in the STA device and conventional case are shown in Fig. 2c. The STA braiding enables the zero mode to transfer from waveguide A to S within this short distance (24 μm, see bottom panel), while the conventional process only leads to partial transfer at the same distance (top panel). Further analysis of the zero-mode occupancy during the pumping process reveals that the conventional braiding path results in significant crosstalk between the pumped zero mode and other modes (black curve, Fig. 2d). In contrast, the STA path maintains the dominance of the zero mode, ensuring that even in the presence of crosstalk, it ultimately returns to the zero-energy level (red curve, Fig. 2d), thus successfully fulfilling the braiding operation.

Figure 3 shows the simulated light evolutions during the STA braiding process. The system's initial states ($z = 0$) are two zero modes occupying waveguides A and B, respectively, i.e., $|\Psi_1(0)\rangle=(0,1,0,0)^T$, $|\Psi_2(0)\rangle=(0,0,1,0)^T$. After the braiding process, both $|\Psi_1\rangle$ and $|\Psi_2\rangle$ undergo energy exchange between waveguides A and B. However, $|\Psi_1\rangle$ follows the A-S-B path without geometric phase inversion (see Fig. 3a, where the input and output have the same phase indicated by the red color), while $|\Psi_2\rangle$ undergoes a single pumping process from B to A (Fig. 3b, where the input and output fields are in red and blue, indicating a phase difference of $\pi$). Therefore, the output geometric phase of $|\Psi_2\rangle$ differs from the initial phase by $\pi$ (the accumulated dynamical phases are both $\beta_0 L$), which is the expected outcome of the Y gate. Following the same STA-braiding principle, the Pauli X and Z gates can also be designed and realized in the tri-layer silicon photonic platform (see **Supplementary Information S2**).

**Experimental observations**

We fabricated the tri-layer-integrated waveguide samples using E-beam lithography with AR-N 7520 resists and an inductively coupled plasma etching process[39]. The bottom layer waveguides are fabricated in a silicon wafer on a sapphire substrate. A layer of SU-8 with a thickness of 530 nm covering the bottom waveguides is prepared, followed by the coating of a second silicon film with a thickness of about 220 nm, which is further processed to form the second layer waveguides. The third layer of silicon is fabricated following the same procedures (see **Methods** for fabrication details).

To clearly observe the fabricated trilayer waveguide structure, we used a focused ion beam (FIB) to cut holes into the surface of the sample, allowing the cross-section of the waveguide sample to be viewed through a scanning electron microscope (SEM), as shown in the bottom panels of Figs 4a-c. Here, Figures 4a, 4b, and 4c illustrate cross-section of the braiding steps 0-1, 1-2, and 2-3, respectively, and the corresponding top panels show the schematics of each cross-section. In optical measurements, a laser light ($\lambda = 1550$ nm) was focused into the waveguide lattice through an input grating coupler. The transmitted light, scattered from the extended output ports, can be collected for analysis. Both the coupling-in and coupling-out processes were captured using a near-infrared charge-coupled device (CCD) camera.

Figures 4d and 4e demonstrate that the two zero modes can indeed switch the light dwelling (A-B and B-A), as required by the braiding process of the Y gate. To further confirm the geometric π phase difference of the two outcomes, two identical Y-gate braiding structures (e.g., $Y_1$ and $Y_2$) were excited with different input ports (e.g., $A_1B_2$, which means inputs from waveguide A in the $Y_1$ device and waveguide B in the $Y_2$ device) or the same ports (e.g., $A_1A_2$), and their outputs were combined to interfere with each other. As shown in Fig. 4f, the experimental observations confirm that outputs from different input waveguides ($A_1B_2$) had opposite phases (no light comes from the central port), while outputs from the same input waveguides ($A_1A_2$) had identical phases (the intensity of the central output is observed to be enhanced), confirming that the braiding device achieved the π geometric phase difference of the Y gate.

**STA non-Abelian braiding**

The successful demonstration of the Y gate shows that our STA design and the integrated photonic platform can be extended to obtain the non-Abelian braiding of (*N*) modes. As a proof of concept, we realized the three-mode non-Abelian braiding operations, representing the minimalist and simplest non-Abelian braiding process described by a braid group ($B_3$), as shown in the schematics in Figs 5a and 5c. The three-mode STA non-Abelian braiding structure comprises seven waveguides. The three primary waveguides, labeled A, B, and C, are positioned sufficiently far from one another, necessitating coupling through auxiliary waveguides $X_1$ and $X_2$, which are individually cut into three parts similar to the two-mode braiding process.

The Hamiltonian of the system supports three degenerate zero modes that constitute the braiding subspace. In this case, exchanging $|\Psi_2\rangle$ and $|\Psi_1\rangle$ is captured by the matrix $G_1 = \begin{pmatrix} Y & 0 \\ 0 & 1 \end{pmatrix}$ and the exchange of $|\Psi_3\rangle$ and $|\Psi_2\rangle$ is given by $G_2 = \begin{pmatrix} 1 & 0 \\ 0 & Y \end{pmatrix}$. Here, $|\Psi_i\rangle$ represents a state vector with its three elements indicating the wavefunction in waveguides A, B, and C, respectively. It is easy to check that $G_1G_2 \neq G_2G_1$, which is a fundamental characteristic that classifies $B_3$ as a non-Abelian group. The simulation results in Figs. 5b and 5d show the evolution of the light field during the three-mode braiding process. When the braiding sequence is $G_2G_1$, the output light signal sequentially shifts one site to the right. Conversely, the output sequentially shifts one site to the left with the $G_1G_2$ sequence. The different braiding sequences result in opposite outcomes, verifying the non-Abelian nature of the braiding process. Notably, the STA non-Abelian braiding process is accomplished within 144 μm, nearly three orders of magnitude shorter than the state-of-the-art laser-written photonic waveguide systems, which operate on a millimeter scale[22]. It is also ~3.8 times smaller compared to the conventional design on the same silicon platform. This miniaturization could enhance the integration density of non-Abelian photonic chips and minimize the impact of dimensional uncertainties due to nanofabrication, thereby enabling scaling to larger circuits with more complex non-Abelian photonic device networks.

**Discussion**

In summary, we have, for the first time, achieved STA braiding of photonic modes within 3D silicon photonic chips. Utilizing a trilayer waveguide configuration and applying the STA strategy to the hopping amplitudes, we successfully executed fast zero-mode braiding operations functioning as Pauli X, Y, and Z gates. Further experimental observation of non-Abelian braiding at telecommunication wavelengths underscores the practical applicability of the STA braiding scheme. Compared to previous works, our results demonstrate the feasibility of realizing non-Abelian devices with significantly reduced sizes, facilitating the construction of quantum logic gates in integrated photonic chips and showcasing the potential to implement more complex non-Abelian optical operations for manipulating photons and light at micro- and nanoscale.

**Methods**

**Fabrication of the trilayer silicon waveguide samples**

The waveguide arrays and grating nanostructures are fabricated using the method of electron-beam lithography and ICP etching process. The substrate used herein is 220 nm silicon deposition on an alumina substrate, and the substrates are cleaned in ultrasound bath in acetone and DI water for 10 min respectively and dried under clean nitrogen flow. Then a layer of MA-N2405 photoresist film is spin-coated onto the substrate and baked at 90℃ for 3 min. After that, the sample is exposed to electron beam in E-beam writer (Elionix, ELS-F125) and developed to form the MA-N2405 nanostructures. Then, the sample is transferred into HSE Series Plasma Etcher 200 and etched with $C_4F_8$ and $SF_6$ (the flow rates of these two types of gases are 75 sccm:30 sccm). After the ICP etching, the remaining MA-N2405 is removed by using an $O_2$ plasma for 5 min. Before the fabrication of the second layer of nanostructures, 530 nm SU-8 resist is spin-coated onto the sample and baked at 200℃ for 30 min for protection.

Then another α-Si layer was deposited on the SU-8 using the plasma-enhanced chemical vapor deposition to a final thickness of 220 nm. Repeat the above process twice to fabricate the second and third layers afterwards (see **Supplementary Fig. S3** for the fabrication flow).

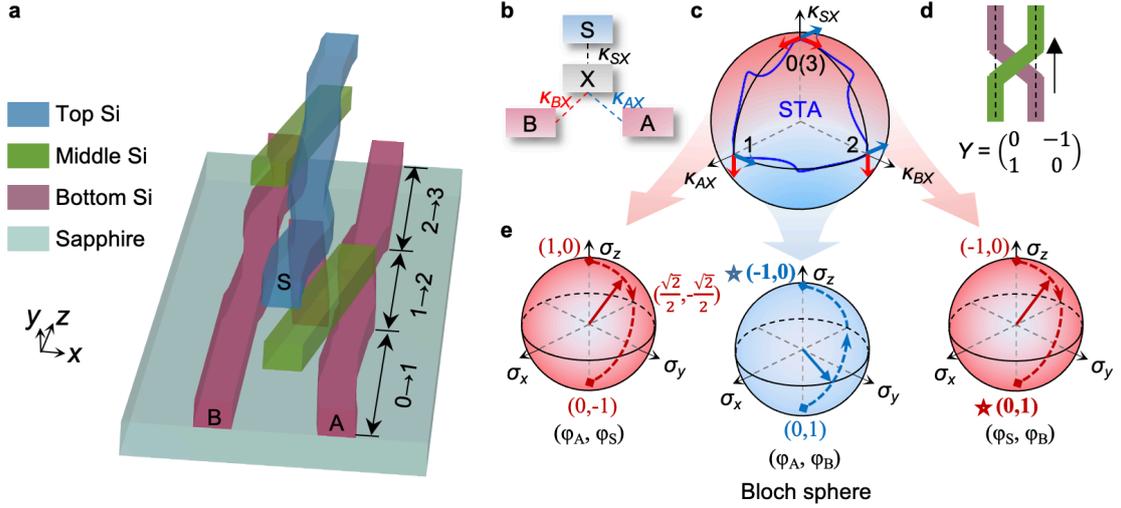

**Fig. 1 Braiding on trilayer silicon photonic chips. a** A schematic of the braiding structure consisting of four waveguides implemented in tri-layered silicon photonics platform. **b** The cross-section of the setup in tight-binding model, including coupling between waveguide-X to waveguide-A, -B, and –S, respectively. **c** The system sustains two degenerate zero modes, whose parallel transport on the 2-sphere exchanges the two modes. The blue trajectory indicates the STA braiding path in the parameter spaces. This process realizes the Y gate, whose braiding diagram is shown in (**d**). **e** The zero modes can be mapped onto Bloch spheres of different subspaces for three braiding steps, respectively. The red and blue dashed paths indicate the zero mode evolutions on the Bloch sphere as performing the braiding operations.

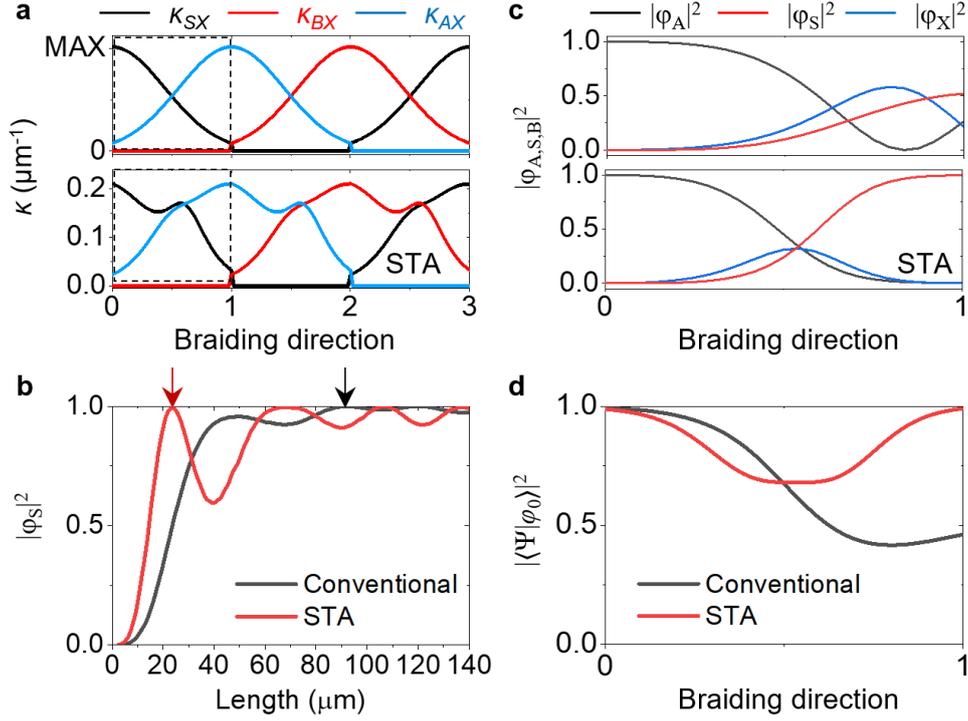

**Fig. 2 Design of STA braiding. a** The original modulation profiles of the coupling coefficients (top). The coupling coefficients in STA braiding process (bottom) at the working wavelength $\lambda=1550$ nm. **b** The transferred light intensity $|\varphi_S|^2$ as a function of the length of the subsystem (waveguide-A, X, and S). The STA braiding allows complete energy transfer within a short system length 24 μm (marked by red arrow), while the conventional process requires at least 92 μm to allow complete transfer (black arrow). **c** Evolution of the light intensity in the subsystem along the braiding direction (*step 0-1*) during the conventional (top) and STA (bottom) process. **d** Zero-mode occupancy $|\langle\Psi|\varphi_0\rangle|^2$ during the braiding process. The evolution length is 24 μm for (**c**) and (**d**).

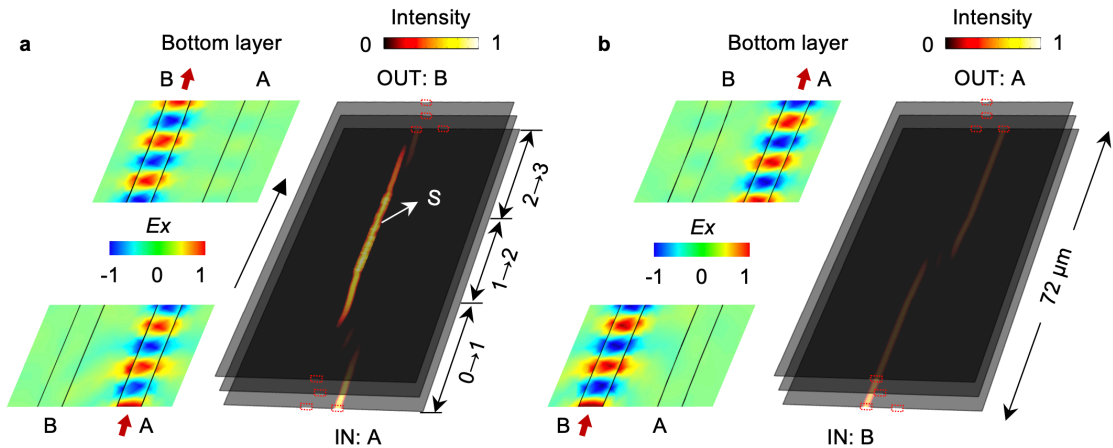

**Fig. 3 Simulation results of STA braiding in tri-layer silicon waveguides. a** For input from waveguide-A. Left panel: Evolution of $E_x$ field at the input and output ends. Right panel: Evolution of light intensity in the STA braiding process. The red boxes in the input and output ends mark the position of the waveguides. **b** Corresponding results for waveguide-B input.

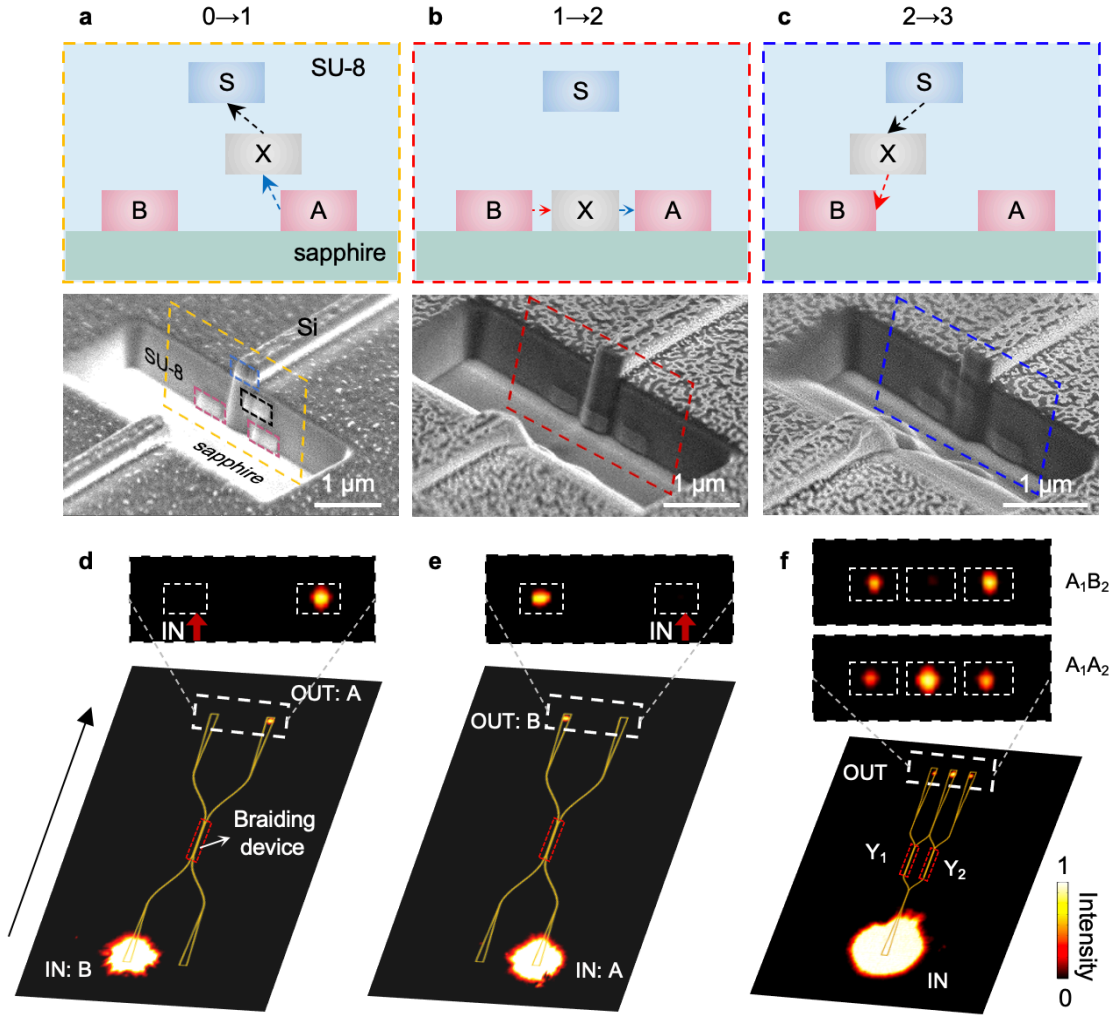

**Fig. 4 Experimental results. a-c** Schematics for different stages of the trilayer waveguide systems (top) and scanning electron microscope (SEM) images for fabricated samples (bottom). **d-f** CCD recorded light couple-in and –out processes in the braiding samples of (**d**) waveguide-B input, (**e**) waveguide-A input, and (**f**) double Y-gate interference results. The top panels show the experimentally captured zoom-in pictures of the output ports. In (**f**), two identical Y-gate structures ($Y_1$, $Y_2$) were input simultaneously, and the interference of their output states reveals the geometric phase. Specifically, the upper output result corresponds to $A_1B_2$ inputs (i.e., inputs from waveguide-A in $Y_1$ device and waveguide-B in $Y_2$ device); while the lower output result corresponds to $A_1A_2$ inputs.

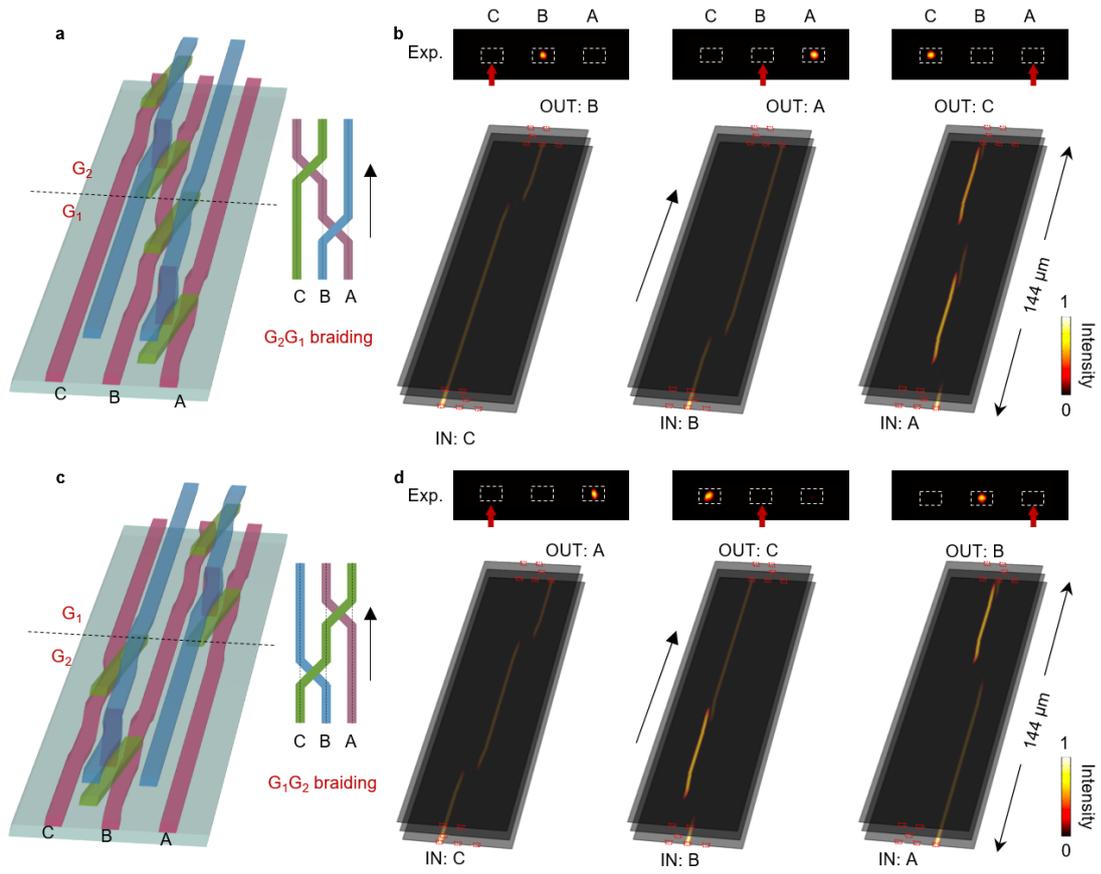

**Fig. 5 Experimental observation of STA non-Abelian braiding. a** A schematic of the $G_2G_1$ braiding structure. **b** Simulation (bottom) and experimental (top) results of the $G_2G_1$ braiding, showing propagation (simulation) and output light field distribution (experiment) from waveguides A, B, and C inputs. The red boxes in the input and output end mark the position of the waveguides. **c,d** Corresponding results of the $G_1G_2$ braiding.